\documentclass[fleqn,usenatbib]{mnras}

\usepackage[T1]{fontenc}

\DeclareRobustCommand{\VAN}[3]{#2}
\let\VANthebibliography\thebibliography
\def\thebibliography{\DeclareRobustCommand{\VAN}[3]{##3}\VANthebibliography}

\usepackage{graphicx}	% Including figure files
\usepackage{amsmath}	% Advanced maths commands
\usepackage{amssymb}	% Extra maths symbols

\usepackage{newtxtext,newtxmath}

\title[Peter Pan discs across stellar mass]{Exploring the possibility of Peter Pan discs across stellar mass}

\author[M. J. C. Wilhelm \& S. Portegies Zwart]{
Martijn J. C. Wilhelm,$^{1}$\thanks{E-mail: wilhelm@strw.leidenuniv.nl}
Simon Portegies Zwart,$^{1}$
\\
$^{1}$Leiden Observatory, Leiden University, PO Box 9513, NL-2300 RA Leiden, the Netherlands
}

\date{Accepted XXX. Received YYY; in original form ZZZ}

\pubyear{2021}

\begin{document}
\label{firstpage}
\pagerange{\pageref{firstpage}--\pageref{lastpage}}
\maketitle

\begin{abstract}
Recently, several accreting M dwarf stars have been discovered with ages far exceeding the typical protoplanetary disc lifetime. These `Peter Pan discs' can be explained as primordial discs that evolve in a low-radiation environment. The persistently low masses of the host stars raise the question whether primordial discs can survive up to these ages around stars of higher mass. In this work we explore the way in which different mass loss processes in protoplanetary discs limit their maximum lifetimes, and how this depends on host star mass. We find that stars with masses $\lesssim$ 0.6 M$_\odot$ can retain primordial discs for $\sim$50 Myr. At stellar masses $\gtrsim$ 0.8 M$_\odot$, the maximum disc lifetime decreases strongly to below 50 Myr due to relatively more efficient accretion and photoevaporation by the host star. Lifetimes up to 15 Myr are still possible for all host star masses up to $\sim$2 M$_\odot$. For host star masses between 0.6 and 0.8 M$_\odot$, accretion ceases and an inner gap forms before 50 Myr in our models. Observations suggest that such a configuration is rapidly dispersed. We conclude that Peter Pan discs can only occur around M dwarf stars.
\end{abstract}


%%%%%%%%%%%%%%%%%%%%%%%%%%%%%%%%%%%%%%%%%%%%%%%%%%

%%%%%%%%%%%%%%%%% BODY OF PAPER %%%%%%%%%%%%%%%%%%

\section{Introduction}

Within the last thirty years the number of planets known to astronomy has increased by many a hundredfold. As of the writing of this piece more than 4300 planets around stars other than our Sun have been confirmed \footnote{\url{https://exoplanets.nasa.gov/discovery/exoplanet-catalog/}}\footnote{\url{http://exoplanet.eu/}, \citet{Schneider2011}}. Current theories of planet formation propose that these planets form in rotationally supported discs around young stars, in this context referred to as protoplanetary discs \citep{Williams2011}. When exactly planets form within these discs is an area of active research, but recent results hint at early formation (within $\sim$0.5 Myr of the formation of the star; e.g. \citet{Tychoniec2020}). An upper limit on this formation time is provided by the lifetime of the disc. Observations of the fraction of stars with protoplanetary discs in different young star forming regions indicate a mean lifetime of about 2-5 Myr, and a maximum lifetime of 10-20 Myr \citep{Ribas2014,Pecaut2016,Richert2018}. Simulations of young star forming regions imply that these limits can be explained by mass loss due to winds driven by far ultraviolet (FUV) radiation from nearby massive stars \citep{Winter2019,Nicholson2019,Concha2019,Concha2021}. 

\citet{Silverberg2016} described the discovery of a circumstellar disc around an M dwarf in the $\sim$45 Myr Carina association, which exhibited a large fractional IR luminosity more typical of a young primordial disc than a debris disc. \citet{Murphy18} then observed the star in the optical, confirming its membership of the Carina association and detecting H$\alpha$ emission consistent with accretion. The latter implied the presence of a gaseous disc rather than a debris disc. Later studies by \citet{Silverberg2020} and \citet{Lee2020} discovered a total of six more accreting M dwarf stars of ages $\sim$50 Myr (plus a brown dwarf of similar age possibly accreting from a disc).

These stars are more than an order of magnitude older than the typical lifetime of protoplanetary discs. This apparent refusal to `grow up' led \citet{Silverberg2020} to coin the term `Peter Pan disc'. In that paper they proposed a number of theories on their origin, but a long-lived primordial protoplanetary disc was most favoured. This possibility was further investigated by \citet{Coleman2020} (hereafter CH20). They found that discs around M dwarfs with plausible initial masses and radii could survive for 50 Myr in conditions where the FUV radiation field was low. In their work, the question was raised why these Peter Pan discs have only been found around low-mass stars. This can a statistical effect, caused by the stellar initial mass function's preference for low-mass stars. Alternatively, a physical process might be at work that disperses discs around higher mass stars more efficiently. It is unlikely to (fully) be a selection effect, because the initial Peter Pan disc candidate selection was done using the Disk Detective citizen science project \citep{Kuchner2016}, which is based on sources with an infrared excess in the {\it WISE} all-sky mid-infrared survey \citep{Wright2010}. Their source selection was not intentionally biased towards particular stellar types, although subtle biases likely exist. 

In this work we investigate whether Peter Pan discs around stars of higher mass (up to $\sim$2 M$_\odot$) can be explained within our current understanding of the evolution of protoplanetary discs. We consider a number of processes that deplete the gas reservoir of protoplanetary discs and how they scale with host star mass. Specifically, we consider the effect of evaporation by internal and external radiation, and of accretion, on the discs' lifetime.

\section{Mass loss processes} \label{sec:mass_loss}

In this section we introduce a number of processes that remove mass from a protoplanetary disc. We also consider how they limit the discs' lifetime, and how this limit depends on host star mass.

We estimate a disc's lifetime $\tau$ under a certain mass loss process from its initial mass $\textrm{M}_\textrm{d}$ and its mass loss rate $\dot{\textrm{M}}_\textrm{d}$ using the simple expression $\tau = \textrm{M}_\textrm{d}/\dot{\textrm{M}}_\textrm{d}$. If both can be described as power law functions of the host star mass ($\textrm{M}_\textrm{d} \sim \textrm{M}_*^\alpha$ and $\dot{\textrm{M}}_\textrm{d} \sim \textrm{M}_*^\beta$), the behaviour of the disc lifetime will depend on the difference of the indices. If $\alpha - \beta < 0$, discs around higher mass stars will have shorter lifetimes. For example, if the disc mass is a constant fraction of the host star mass ($\alpha=1$), and a mass loss process scales quadratically with host star mass ($\beta=2$), the disc lifetime will be inversely proportional to the host star mass. 

In practice, multiple processes will be at work, and the scaling with host star mass will depend on which process is dominant. Below we discuss the three main mass loss processes.

\begin{figure}
    \centering
    \includegraphics[width=1\linewidth]{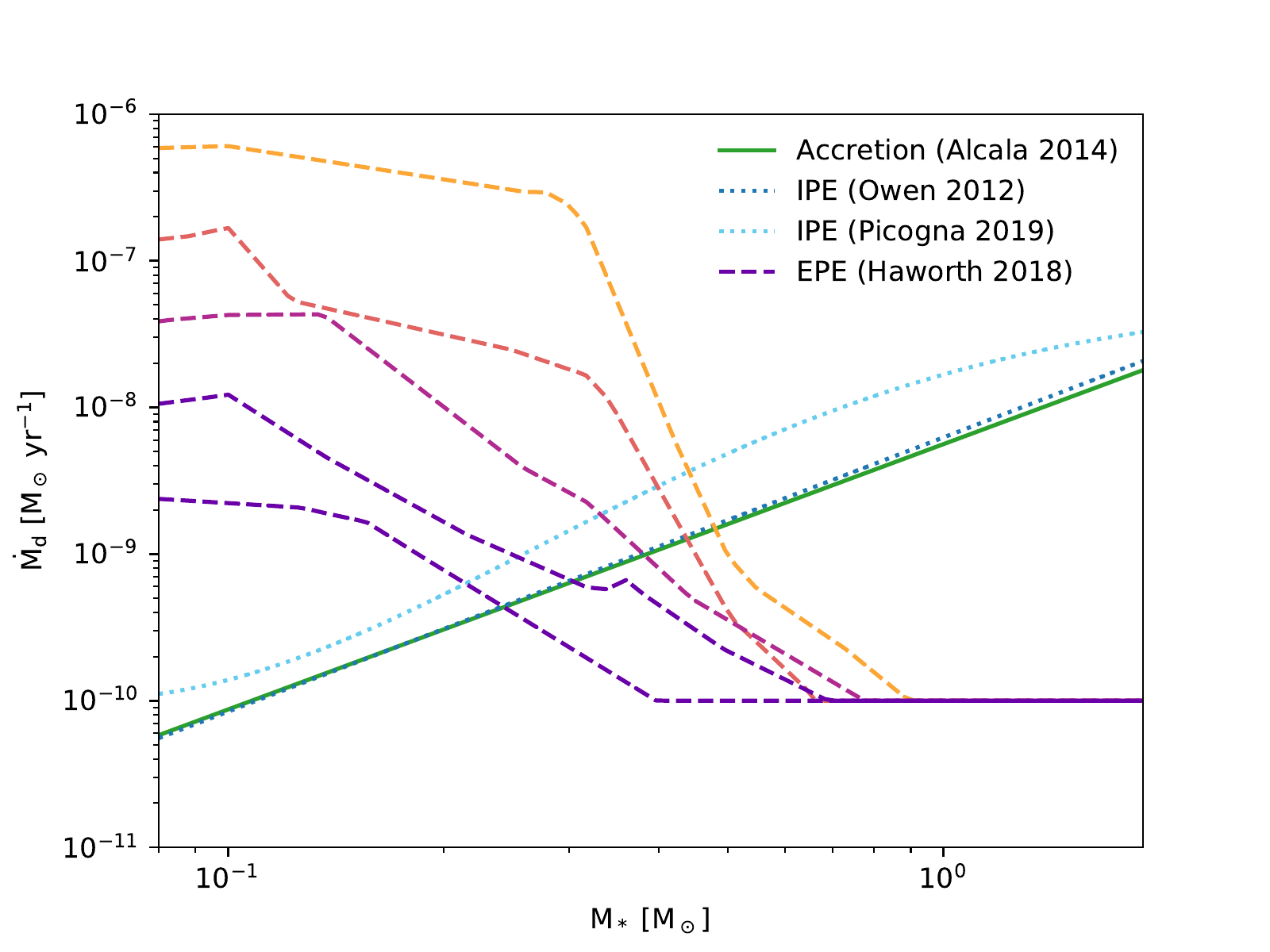}
    \caption{Disc mass loss rates of a number of processes as a function of host star mass. The processes are external photoevaporation (EPE), internal photoevaporation (IPE), and accretion. The dashed lines show the EPE rate as obtained from interpolating on the FRIED grid \citep{Haworth2018}, for a range of disc masses. These are spaced logarithmically between $1.57\cdot 10^{-2}$ M$_{\textrm{Jup}}$ to 2.62 M$_{\textrm{Jup}}$ (from bottom to top). The disc radius is 100 au, and the radiation field is 10 G$_0$. The dotted blue and cyan lines show the IPE rates from \citet{Owen2012} and \citet{Picogna2019}, respectively. The solid green line shows the fitted accretion rate dependence on stellar mass from \citet{Alcala2014}.}
    \label{fig:mdot}
\end{figure}

\subsection{External photoevaporation} \label{sec:epe}

Radiation of stars near a disc's host star can drive a thermal wind from the disc by heating material to the point that it becomes gravitationally unbound. This process is called external photoevaporation (hereafter EPE). It is typically dominated by FUV radiation, although extreme ultraviolet (EUV) radiation dominates close to massive stars \citep{Johnstone1998}. \citet{Adams2004} computed mass loss rates due to external FUV radiation for different disc parameters, and e.g. \citet{Anderson2013} coupled mass loss rates to a dynamic viscous disc model. 

\citet{Haworth2018} produced the FRIED grid, which contains mass loss rates due to EPE by FUV radiation for an extensive grid of host star mass, external radiation field, disc radius, and disc mass. Fig. \ref{fig:mdot} shows, as dashed lines, the linearly interpolated mass loss rate as a function of host star mass for five disc masses. These disc masses are logarithmically spaced from $1.57\cdot 10^{-2}$ M$_{\textrm{Jup}}$ to 2.62 M$_{\textrm{Jup}}$ (from bottom to top). These disc masses fall within the convex hull of the FRIED grid over the full range of stellar masses, ensuring that all plotted values result from interpolation on the grid. The radiation field was chosen to be 10 G$_0$\footnote{G$_0$ is the Habing field \citep{Habing1968}, $1.6\cdot 10^{-3}$ erg s$^{-1}$ cm$^{-2}$. The typical interstellar FUV flux is 1.7 G$_0$ \cite{Draine1978}.}, and the disc radius 100 au. 

As a general trend, the mass loss rate decreases with increasing host star mass and decreasing disc mass. The FRIED grid has a minimum mass loss rate of $10^{-10}$ M$_\odot$ yr$^{-1}$, which is reached at large host star masses for every disc mass. 

Note that CH20 used a simplified expression for the mass loss rate. Instead of relating it to the strength of the FUV field, they fixed a reference mass loss rate, which was then scaled linearly with the disc radius. This was because the lowest radiation field in the FRIED grid (10 G$_0$) was expected to be too high to result in Peter Pan discs. Indeed, Fig. \ref{fig:mdot} shows that the mass loss rate of discs around M dwarf stars can easily be greater than $10^{-8}$ M$_\odot$ yr$^{-1}$, which CH20 found to be too high to result in Peter Pan discs.

\subsection{Internal photoevaporation} \label{sec:ipe}

The radiation of the disc's host star can also drive a thermal wind, a mass loss process generally termed internal photoevaporation (hereafter IPE). Detailed simulations were performed by e.g. \citet{Owen2012,Picogna2019}. \citet{Owen2012} derived that the resulting mass loss rate is nearly independent of host star mass (a power law index of -0.068). However, this mass loss rate does depend strongly on the X-ray luminosity (a power law index of 1.14). The luminosity in turn depends on stellar mass. \citet{Flaccomio2012} found that the characteristic X-ray luminosity of T Tauri stars depends on stellar mass with a power law index of 1.7. Taken together, this results in a power law index $\beta$ of 1.87.

\citet{Picogna2019} later improved the modelling of \citet{Owen2012} and found a more sigmoidal behaviour of the mass loss rate as a function of X-ray luminosity. This relation becomes nearly constant at X-ray luminosities corresponding to $\sim$1 M$_\odot$, but below that has a slope similar to \citet{Owen2012}'s relation.

In Fig. \ref{fig:mdot} we show the IPE rates of \citet{Owen2012} and \citet{Picogna2019} as a function of host star mass, assuming the characteristic X-ray luminosity results of \cite{Flaccomio2012}, as dotted lines.

The X-ray luminosities used to derive the IPE rates are based on observations of a population of T Tauri stars of $\sim$1 Myr, and stars evolve onto the main sequence on timescales of a few to a few tens of Myrs. \citet{Preibisch2005} compared the X-ray luminosities of low-mass stellar populations of different ages. They found a decrease of an order of magnitude in luminosity between stars in the Orion Nebula Cluster (the sample our luminosity function is based on) and stars in the Pleiades cluster (estimated to be $\sim$110 Myr \citep{Dahm2015,Gossage2018}. 

\citet{Gregory2016} connect the decrease in X-ray luminosity in pre-main sequence stars with the development of a radiative core, which happens for stars more massive than 0.35 M$_\odot$. They find that after this transition the X-ray luminosity decreases with time approximately as $L_X \propto t^{-2/5}$, broadly consistent with an order of magnitude decrease from 1 to 110 Myr. The results from \citet{Johnstone2020} show that this decrease also holds for stars below 0.35 M$_\odot$. 

Because this decrease in X-ray luminosity can be considerable over the lifetime of a Peter Pan disc we adopt the following luminosity evolution:

\begin{equation} \label{eq:xray_decrease}
    L_X\left(t\right) = \begin{cases} 
        L_{X,0} & t < 1 \textrm{ Myr} \\
        L_{X,0} \left(\frac{t}{1\textrm{ Myr}}\right)^{-2/5} & t > 1 \textrm{ Myr},
    \end{cases}
\end{equation}

where $L_{X,0}$ is the characteristic X-ray luminosity relation of \cite{Flaccomio2012}.

\subsection{Accretion} \label{sec:accr}

The rate of gas accretion onto the host star is typically found to scale with host star mass, with a power law index of $\sim$2 (e.g. \citet{Alexander2006}). We adopt the accretion rate's dependence on host star mass recovered by \citet{Alcala2014} for young stellar objects in the Lupus star forming region:

\begin{equation}
    \log \dot{\textrm{M}}_\textrm{accr} = \left(1.81\pm 0.20\right) \log \textrm{M}_* - \left(8.25\pm 0.14\right),
\end{equation}

where $\dot{\textrm{M}}_\textrm{accr}$ has units of $\textrm{M}_\odot \textrm{ yr}^{-1}$ and $\textrm{M}_*$ has units of $\textrm{M}_\odot$. In Fig. \ref{fig:mdot} we show this dependence, without error bars, as a solid green line. This dependence results in a $\beta$ index of $1.81\pm 0.20$, similar to IPE.

\section{Simulations}

\subsection{Disc model} \label{sec:disc_model}

For our simulations we use the disc model of \citet{Concha2021} with a number of additions. This model uses the {\sc vader} code \citep{Krumholz2015} to simulate the viscous evolution of a protoplanetary disc, and includes a module that implements the effects of EPE. This model is set up using the AMUSE framework \citep{PortegiesZwart2009,Pelupessy2013}.

We included IPE using the mass loss rates and profiles from \citet{Picogna2019}. The profiles are scaled with respect to the stellar mass using the scaling relations from \citet{Owen2012}. We use the results from \citet{Flaccomio2012} for the characteristic X-ray luminosity of T Tauri stars as a function of stellar mass in the computation of the mass loss rates. These luminosities are decreased with time following Eq. \ref{eq:xray_decrease}. 

We also included a decrease in the prescribed mass accretion rate with time. According to viscous disc theory \citep{LyndenBell1974}, the accretion rate evolves with time as follows:

\begin{equation} \label{eq:viscous_accr}
    \dot{\textrm{M}}_\textrm{accr}\left(t\right) = \frac{\textrm{M}_\textrm{d}}{2t_s}\left(1 + \frac{t}{t_s}\right)^{-3/2},
\end{equation}

where $t$ is the current time and $t_s$ is the viscous timescale:

\begin{equation}
    t_s = \frac{\textrm{R}_\textrm{d}^2}{3} \frac{\Omega}{\alpha c_s^2}.
\end{equation}

In these expressions, $\textrm{M}_\textrm{d}$ and $\textrm{R}_\textrm{d}$ are the disc's initial mass and radius, respectively, and $c_s$ and $\Omega$ are the sound speed and Keplerian orbital angular frequency at $\textrm{R}_\textrm{d}$. Because we fix the accretion rate we do not use Eq. \ref{eq:viscous_accr} directly, but scale the accretion rate with a factor $\left(1 + t/t_s\right)^{-3/2}$. 

Using our model's midplane temperature relation\footnote{$\textrm{T}\left(\textrm{R}, \textrm{M}_*\right)=100 \textrm{ K}\left(\frac{\textrm{R}}{\textrm{au}}\right)^{-1/2} \left(\frac{\textrm{M}_*}{\textrm{M}_\odot}\right)^{1/4} $} and assuming the ideal gas law, we get the following scaling relation for $t_s$:

\begin{equation} \label{eq:viscous_time}
    t_s = 11 \textrm{ Myr} \left(\frac{\alpha}{10^{-3}}\right)^{-1} \left(\frac{\textrm{R}_\textrm{d}}{200 \textrm{ au}}\right) \left(\frac{\textrm{M}_*}{\textrm{M}_\odot}\right)^{1/4}.
\end{equation}

Note that this is shorter than the typical Peter Pan disc age for relatively high viscosities ($\alpha=10^{-3}$), and somewhat longer for relatively low viscosities ($\alpha=10^{-4}$). In both cases, the accretion rate will decrease considerably over a Peter Pan disc's lifetime.

We recall that in this model, situations can emerge in which the prescribed accretion can not be sustained because the transport rate towards the inner disc edge is too low. If this happens we first try to reduce the accretion rate such that the first cell can be completely drained; if that fails, we switch to the vanishing-torque boundary condition. 

Analogous to CH20 we fix a reference EPE rate for our simulations. This mass loss rate is then scaled linearly with the disc radius. However, as Fig. \ref{fig:mdot} demonstrates that the mass loss rate can depend strongly on the host star mass, we do run simulations using the FRIED grid with a fixed radiation field of 10 G$_0$ (the lowest radiation field on the grid) for comparison.

\subsection{Initial conditions} \label{sec:ics}

We must carefully consider the dependence of the disc mass on host star mass, as the typical disc lifetime scales linearly with the disc mass. The maximum mass of a disc is in part determined by its stability against self-gravity. \citet{Haworth2020}, from simulations, found a relation between the maximum stable disc mass $\textrm{M}_{\textrm{d,max}}$, its radius $\textrm{R}_\textrm{d}$, and its host star's mass $\textrm{M}_*$:

\begin{equation} \label{eq:toomreq}
    \textrm{M}_{\textrm{d,max}} = 0.17 \textrm{ M}_\odot \left(\frac{\textrm{R}_\textrm{d}}{100 \textrm{ au}}\right)^{1/2} \left(\frac{\textrm{M}_*}{\textrm{M}_\odot}\right)^{1/2}.
\end{equation}

This leaves the disc radius as a free parameter. As this is likely not constant as a function of stellar mass, we seek an additional constraint. We use the relation of \citet{Andrews2010} between disc mass and disc radius, where $\textrm{M}_\textrm{d} = 2\cdot 10^{-3} \textrm{ M}_\odot \left(\frac{\textrm{R}_\textrm{d}}{10\textrm{ au}}\right)^{1.6}$. This relation is found for discs in the $\sim$1 Myr old Ophiuchus region; later work by \citet{Hendler2020} found similar slopes in other young star forming regions, and a slightly steeper slope for a $\sim$5 Myr region. As their age is younger than a viscous timescale, we consider this relation to be representative of the initial structure of protoplanetary discs. 

Combining these relations yields the following disc masses and disc radii as a function of stellar mass:

\begin{equation} \label{eq:max_mass}
    \textrm{M}_{\textrm{d,max}} = 0.24 \textrm{ M}_\odot \left(\frac{\textrm{M}_*}{\textrm{M}_\odot}\right)^{0.73},
\end{equation}

\begin{equation}
    \textrm{R}_\textrm{d} = 200 \textrm{ au} \left(\frac{\textrm{M}_*}{\textrm{M}_\odot}\right)^{0.45}.
\end{equation}

These initial conditions represent the maximum mass protoplanetary discs can have while still resembling observed young protoplanetary discs. This is an upper limit, and allows us to determine an upper limit to the disc lifetime.

We note that the initial radius for a 0.1 M$_\odot$ star, as considered by CH20, is 71 au. This is slightly smaller than their fiducial value of 100 au, and corresponds to an initial disc mass 83\% of their fiducial value. 

In our initial analytic estimates of the disc lifetime in Sec. \ref{sec:mass_loss}, we would now have $\alpha=0.73$. The power law index $\beta$ for EPE varies but is generally negative, and consequently leads to increasing disc lifetimes with increasing host star mass. For IPE, the Owen model yields $\beta=1.87$ (with the Picogna model varying but similar), and for accretion, $\beta=1.81$. As a result, if either of these processes is dominant, we expect disc lifetimes to decrease with host star mass. 

Finally, we note that when we use a reference EPE rate, we scale with the initial radius of each individual disc. Together with the fact that the EPE rate also depends on host star mass, this means that a certain reference EPE rate does not correspond to one specific radiation environment.

\subsection{Model grid}

We run series of simulations with different EPE mass loss rates and disc turbulent viscosities. Each series consists of 30 discs, the host star masses of which are logarithmically spaced between 0.08 M$_\odot$ (the traditional hydrogen burning limit) and 1.9 M$_\odot$ (the maximum host star mass in the FRIED grid). We run series with reference EPE rates of 10$^{-8}$, 10$^{-9}$, and 10$^{-10}$ M$_\odot$ yr$^{-1}$, and with rates interpolated on the FRIED grid (though see Section \ref{sec:fried_extrapol} about limited grid coverage), for a radiation field of 10 G$_0$, which is the smallest radiation field available. This to understand how host star mass and radiation field interact in determining the EPE rate. We run series with $\alpha=10^{-3}$ and $\alpha=10^{-4}$, for a total of 8 series. Each is run for 60 Myr. We also summarise these parameters in Table \ref{tab:param_table}.

\begin{table}
    \centering
    \begin{tabular}{c|c|c}
        Parameter & Symbol & Values \\
        \hline
        Host star mass & $\textrm{M}_*$ & $\left[0.08, 1.9\right]\textrm{ M}_\odot$, log spaced \\
        Disc mass & $\textrm{M}_\textrm{d}$ & $0.24 \left(\textrm{M}_*/\textrm{M}_\odot\right)^{0.73}\textrm{ M}_\odot$ \\
        Disc inner radius & - & 0.01 au \\
        Disc outer radius & $\textrm{R}_\textrm{d}$ & $200 \left(\textrm{M}_*/\textrm{M}_\odot\right)^{0.45} \textrm{ au}$ \\
        Disc temperature & $\textrm{T}$ & 100 $\left(\textrm{R}/\textrm{au}\right)^{-1/2} \left(\textrm{M}_*/\textrm{M}_\odot\right)^{1/4} \textrm{ K}$ \\
        Disc viscosity & $\alpha$  & $\{10^{-4}, 10^{-3}\}$ \\
        Reference EPE rate & - & $\{ 10^{-10}, 10^{-9}, 10^{-8} \} \textrm{ M}_\odot \textrm{ yr}^{-1}$\\
        Radiation field & - & $\{10\} \textrm{ G}_0$ \\
        \hline
    \end{tabular}
    \caption{Overview of the parameters used for the models described in this work. The reference EPE rates are scaled linearly with the disc radius during the simulation. During a simulation series, we use either a reference EPE rate, or compute the EPE rate from the radiation field. }
    \label{tab:param_table}
\end{table}

\section{Results}

\subsection{Disc lifetimes}

\begin{figure*}
    \centering
    \includegraphics[width=0.49\linewidth]{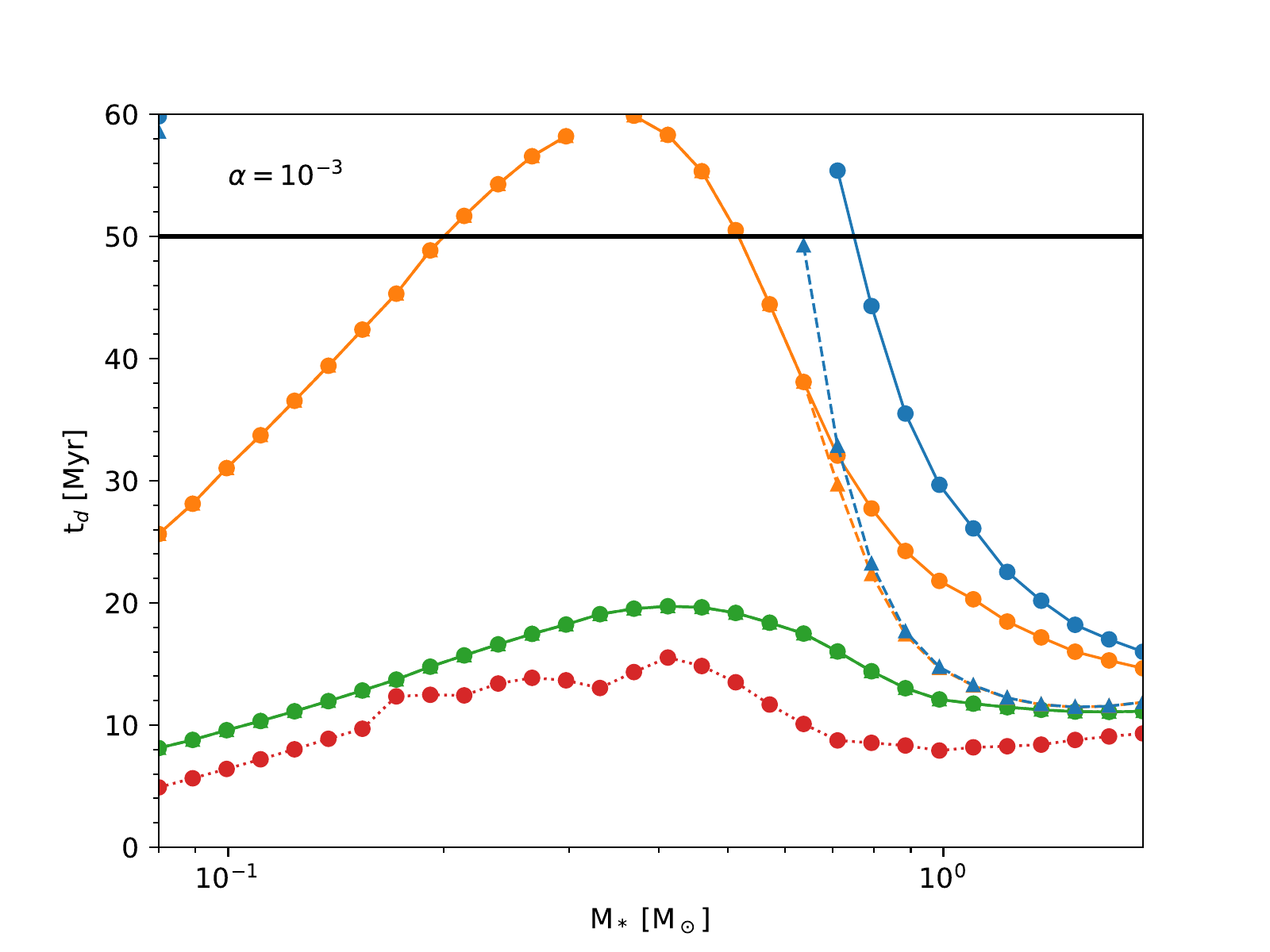}
    \includegraphics[width=0.49\linewidth]{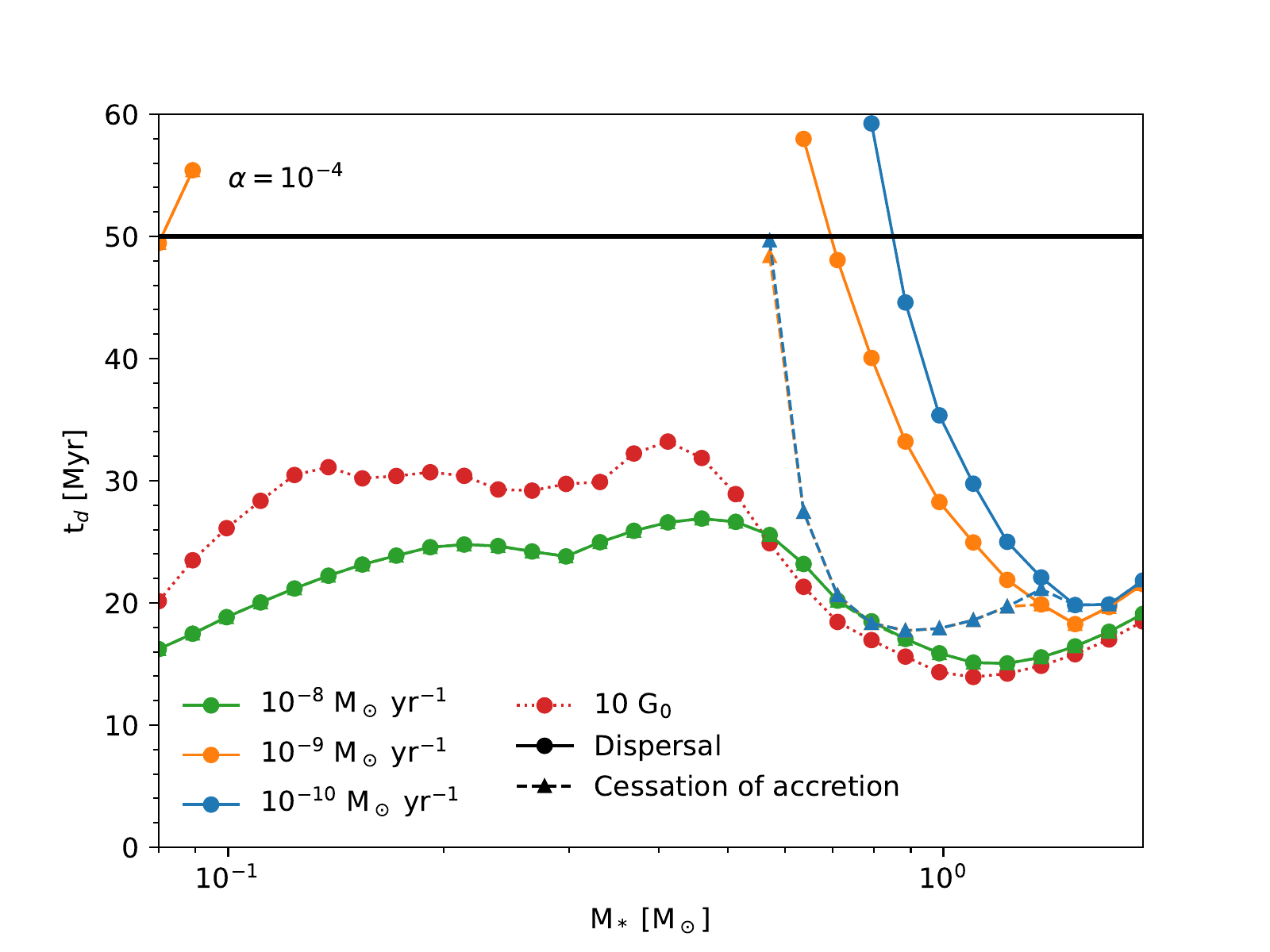}
    \caption{The disc lifetimes (points) and moments of cessation of accretion (triangles; taken to be equal to the moment of dispersal if it has not happened before dispersal) as a function of host star mass for a viscosity parameter of $\alpha=10^{-3}$ (left) and $\alpha=10^{-4}$ (right). Each point corresponds to a simulation, connecting lines are added to help guide the eye. Solid lines correspond to the dispersal of discs under a 'fixed' EPE rate (these reference rates are scaled linearly with the disc outer radius as the disc evolves). The dotted line corresponds to EPE rates as computed from the FRIED grid, with a radiation field of 10 G$_0$. The dashed lines correspond to the cessation of accretion (which does not happen before dispersal for EPE rates of $10^{-8}$ M$_\odot$ yr$^{-1}$ and a radiation field of 10 G$_0$). The horizontal black solid line corresponds to 50 Myr, the characteristic age of Peter Pan discs.}
    \label{fig:disp_times}
\end{figure*}

The lifetimes of our model discs depend on their host star mass, EPE rate, and turbulent $\alpha$ viscosity. We illustrate these dependencies in Fig. \ref{fig:disp_times}.

For relatively high viscosities ($\alpha=10^{-3}$), discs subject to EPE rates of $10^{-10}$ M$_\odot$ yr$^{-1}$ reach the observed age of Peter Pan discs of about 50 Myr for host star masses $<0.7$ M$_\odot$. At higher host star masses, the disc lifetimes decrease steeply, to 20 Myr at 1.4 M$_\odot$. 

At an EPE rate of $10^{-9}$ M$_\odot$ yr$^{-1}$, 50 Myr is reached for discs with host masses between 0.2 and 0.5 M$_\odot$. The increasing trend in disc lifetime with stellar mass below that range is due to EPE being the dominant mass loss process, as discussed in Section \ref{sec:epe}. Similarly, the decreasing trend above that range is due to IPE and accretion being the dominant mass loss processes. The peak is at about 0.3 M$_\odot$, where the combined IPE and accretion rates become larger than the reference EPE rate (see Fig. \ref{fig:mdot}). 

The disc lifetimes corresponding to a radiation field of 10 G$_0$ are a few Myr shorter than those corresponding to a reference EPE rate of $10^{-8}$ M$_\odot$ yr$^{-1}$, but follow the same general trend. Compared to lower EPE rates, the peaks of the lifetime distributions have moved to slightly larger host star masses as the IPE and accretion rates need to be larger to match the EPE rate. 

For smaller viscosities ($\alpha=10^{-4}$), the maximum host star mass for which the 50 Myr threshold is reached has increased to 0.9 M$_\odot$. For an EPE rate of $10^{-9}$ M$_\odot$ yr$^{-1}$ this limit increases to 0.7 M$_\odot$. For the series with an EPE rate of $10^{-8}$ M$_\odot$ and a radiation field of 10 G$_0$ the lifetimes have also increased compared to those with a higher viscosity, but are still limited to below 35 Myr. 

We again see that the disc lifetimes corresponding to a radiation field of 10 G$_0$ are similar to those corresponding to a reference EPE rate of $10^{-8}$ M$_\odot$ yr$^{-1}$, although they diverge at small host star masses. This implies that the EPE rate across stellar mass, for the initial radii corresponding to those stellar masses, is almost constant for a radiation field of 10 G$_0$. 

For a viscosity of $10^{-4}$ there is an upturn in disc lifetime above 1 M$_\odot$. This is caused by the flattening of the IPE rate at those masses (see Fig. \ref{fig:mdot}), leading to $\alpha-\beta$ switching sign. This upturn is not present at higher viscosity because EPE is more efficient there.

The results described above imply that discs can potentially reach ages even older than the 60 Myr that we run our simulation for. In order to explore the limits, we ran the series with $\alpha=10^{-4}$ and an EPE rate of $10^{-10}$ M$_\odot$ yr$^{-1}$ for 100 Myr. We found that the disc around the 0.08 M$_\odot$ star reached an age of 91 Myr, and the one around the 0.71 M$_\odot$ star reached an age of 79 Myr. All discs with host star masses between these values were not dispersed by 100 Myr.

\subsection{Accretion rates} \label{sec:results_accr_rate}

\begin{figure*}
    \centering
    \includegraphics[width=0.49\linewidth]{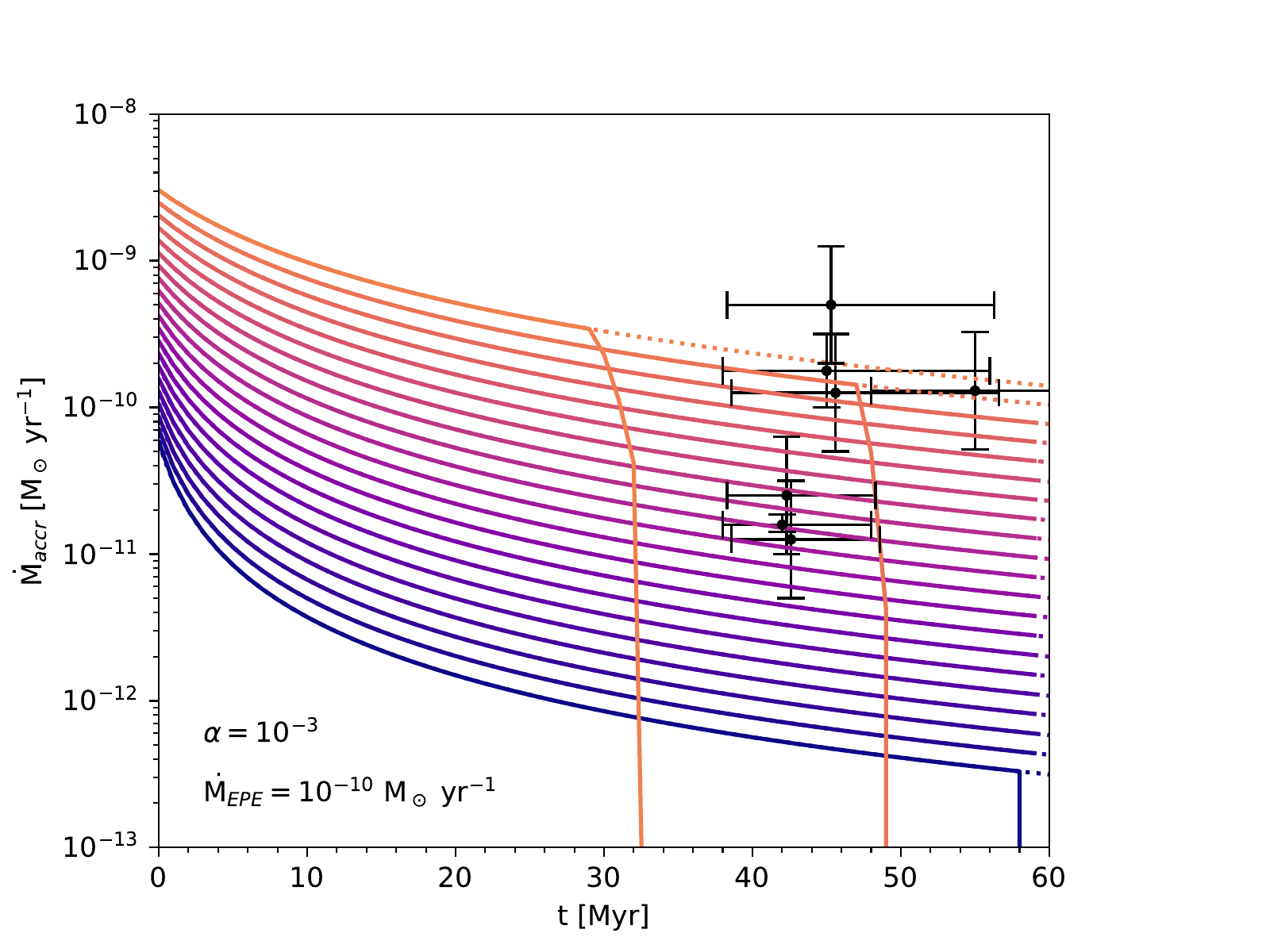}
    \includegraphics[width=0.49\linewidth]{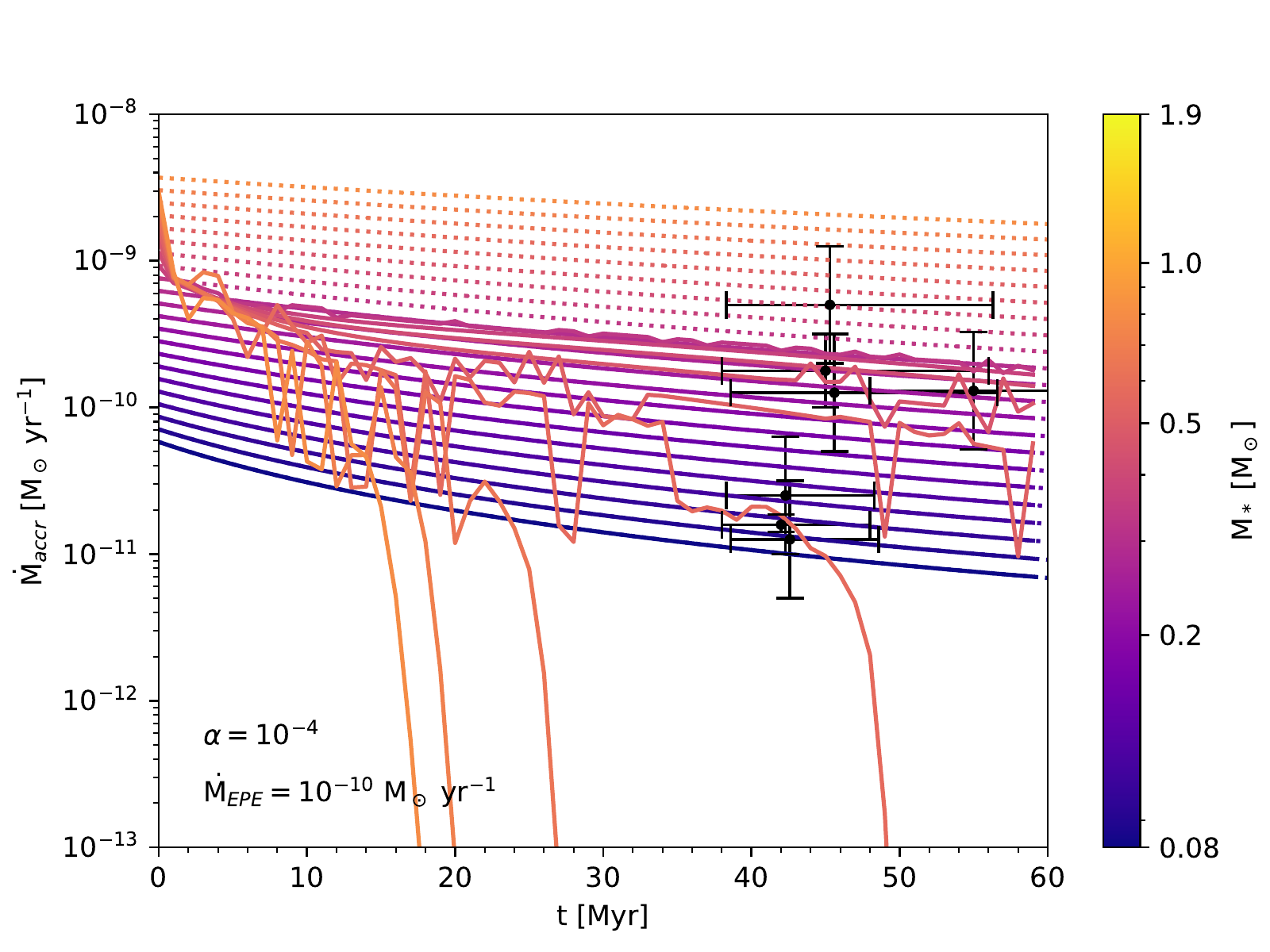}
    \caption{The accretion rates as a function of time for all discs that live to at least 50 Myr under an EPE rate of $10^{-10}$ M$_\odot$ yr$^{-1}$, for a viscosity parameter of $\alpha=10^{-3}$ (left) and $\alpha=10^{-4}$ (right). The dotted lines are the prescribed accretion rates, and the solid lines are the actual accretion rates. The colour indicates host star mass, with dark purple corresponding to 0.08 M$_\odot$. Note that the accretion rate is a monotically increasing function of host star mass. Also shown are the observed ages and accretion rates of the Peter Pan discs listed in \citet{Silverberg2020} and \citet{Lee2020} as black points with error bars (the latter did not provide error estimates, so we opted to conservatively display the largest error bars of the sample of \citet{Silverberg2020}). The two clusters of three points have the same ages, but we have displaced them horizontally for visibility. The left-most point of the clusters corresponds to their true age.}
    \label{fig:accr_rate}
\end{figure*}

Fig. \ref{fig:accr_rate} shows, as a function of time, the mass accretion rates of all discs that live to at least 50 Myr, for $\alpha$ viscosities of $10^{-3}$ and $10^{-4}$. The accretion rates decline with time, as we discussed in Section \ref{sec:disc_model}. This decline is faster for discs with higher viscosity. For a number of discs, the accretion rate drops to 0 before 50 Myr. This means that the host star is somehow stopped from accreting even though a considerable amount of material is still present. Because initial accretion rate increases with host star mass, we can see that it is the higher mass host stars that cease accretion, and this cessation also occurs earlier for higher mass host stars. A noticeable exception is the lowest mass host star at viscosity $10^{-3}$, which ceases accretion at 58 Myr (which is also dispersed just before 60 Myr). This implies that the distribution of the moment accretion ceases is not a strictly decreasing function of stellar mass but peaked like the disc lifetime distribution. 

We compare the accretion rates from our model to those of the Peter Pan discs described by \citet{Silverberg2020} and \citet{Lee2020}. Note that our model prescribes mass accretion rates, therefore reproducing these values is not our aim; in that regard CH20's is a more accurate model. However, in order to maintain this accretion rate the disc has to be able to supply the material to the inner disc edge, which some discs (especially at lower viscosity) apparently fail to do. Notably, those discs that are able to maintain accretion have accretion rates consistent with those observed (albeit marginally for the highest accretion rate Peter Pan disc). 

In Fig. \ref{fig:disp_times}, we also show the moments discs cease accretion with triangle symbols, if they do so before dispersal. For a given host star mass and viscosity and different EPE rates, this happens at almost the same moment (if the disc is not already dispersed, of course). As a general trend, accretion ceases earlier in discs around more massive stars. For reference EPE rates of $10^{-8}$ M$_\odot$ yr$^{-1}$ and radiation fields of 10 G$_0$, the disc is dispersed before accretion ceases for all host star masses.

\section{Discussion}

We have run a grid of protoplanetary grid models with varying host star masses, turbulent viscosities, and EPE rates. Our models take into account viscous evolution and accretion, EPE and IPE. The initial conditions of our models correspond to the maximum mass and radius for which they are gravitationally stable, and resemble observed young protoplanetary discs. From these models we have obtained the maximum plausible lifetime of protoplanetary discs as a function of host star mass. We have seen that discs can survive up to 50 Myr for a range of host star masses, but that at relatively high masses (this depends on viscosity, but is slightly below solar mass) the maximum lifetimes sharply decrease. For these masses, the maximum lifetime is limited by accretion and IPE, both processes associated with the host star.

\subsection{Maximum disc mass} \label{sec:max_disc_mass}

Our results depend strongly on our assumption of the maximum disc mass as the disc lifetime scales linearly with disc mass. Our assumption in turn depends on the limit for which a disc is gravitationally stable and on the typical mass-radius relation of protoplanetary discs in a 1 Myr old star forming region. 

The gravitational stability is a hard limit on the maximum scale of the disc, although it is calibrated on simulations. This calibration constant is the 0.17 M$_\odot$ in Eq. \ref{eq:toomreq}. \citet{Haworth2020} quote an upper limit of 0.3 M$_\odot$, corresponding to a disc that is optically thin to its host star's radiation. This would increase the maximum disc mass and maximum disc lifetime by a factor of $\sim$2.3 for all host star masses. Consequently, the host star mass threshold for reaching 50 Myr would increase to $\sim$1.2 M$_\odot$. However, considering the high density and large amount of dust in a protoplanetary disc this limit seems unlikely. 

The power law indices of this relation (Eq. \ref{eq:toomreq}) can be derived analytically from considering the Toomre Q stability criterion, and have been confirmed by \citet{Haworth2020}'s more careful modelling. These are unlikely to be subject to change. 

The assumption on the initial disc mass-radius relation is derived from extrapolating observational results of protoplanetary discs in a young star forming region. In this case, increasing the relation's reference disc mass actually decreases the disc mass at constant host star mass. This is because it increases a disc's typical density, making it more gravitationally unstable. The disc must then have a smaller radius to stay stable. 

The observations that this relationship is based on are of a slightly evolved disc population ($\sim$1 Myr). Considering that viscous spreading lowers the typical density of a disc, discs would be more dense at birth than assumed in our initial conditions. As a result their masses at birth would be lower than assumed in this work. Discs would then in general have shorter maximum lifetimes, which consequently lowers the host star mass threshold for reaching 50 Myr.

Concerning the power law index, the results of \citet{Hendler2020} suggest that the slope is more steep for older star forming regions. This implies that the initial slope is, if anything, more shallow than what we used here. This would then yield a steeper slope in Eq. \ref{eq:max_mass}. In turn, this would lead to discs around more massive stars being relatively more long-lived, raising the threshold for reaching 50 Myr to higher masses. 

However, considering that the age of the star forming region this relation is based on is shorter than the typical viscous timescale (see Eq. \ref{eq:viscous_time}), they have likely not evolved much yet. This implies that the relation is representative of the properties of discs at birth.

We note that the results of \citet{Andrews2010} are based on fitting radiative transfer models to the observed dust emission. Although the obtained masses are gas masses, the radii are those of the dust component. \citet{Trapman2019} showed that 1 Myr of dust evolution in a disc can result in an observed gas radius that are a factor of a few higher that the dust radius. With this in mind we can interpret the gas masses obtained by \citet{Andrews2010} as the mass of the gas component contained within the radius of the dust component. Whether the mass-radius relation for the full gas component is the same as for this inner component is uncertain.

\subsection{Cessation of accretion}

\begin{figure}
    \centering
    \includegraphics[width=1\linewidth]{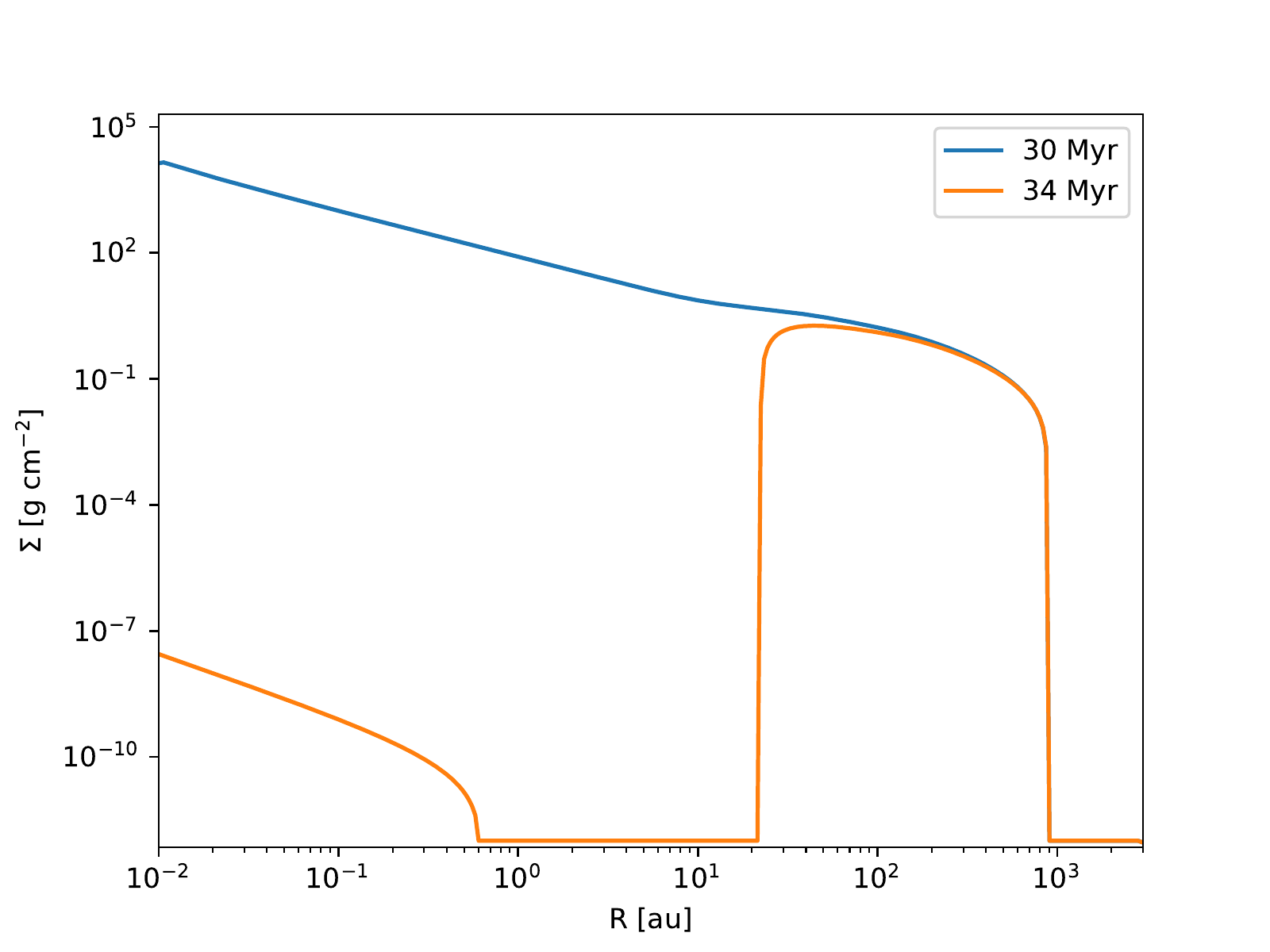}
    \caption{Gas surface density profiles of a protoplanetary disc subject to accretion, IPE, and EPE, at 30 Myr and 34 Myr. The disc has a host star mass of 0.71 M$_\odot$, an $\alpha$ viscosity parameter of $10^{-3}$, and is subject to a reference EPE rate of $10^{-10}$ M$_\odot$ yr$^{-1}$.}
    \label{fig:disk_profiles}
\end{figure}

As discussed in Section \ref{sec:results_accr_rate}, a number of our discs cease accretion before they are dispersed. This behaviour was first theorised by \citet{Clarke2001}. At some point, the IPE rate becomes greater than the accretion rate through the disc, for example through natural evolution or a giant planet choking off accretion \citep{Rosotti2013,Rosotti2015}. The IPE mass loss profile peaks at a few au, and if the mass flow through that region is smaller than the IPE rate, the inner disc can't be replenished. It is eventually accreted onto the host star, creating an inner gap. In Fig. \ref{fig:disk_profiles}, we show the density profiles of a disc at two different times. The disc is the most massive, or top, one from the left panel of Fig. \ref{fig:accr_rate}, which ceases accretion just over 32 Myr. The profiles are from 30 and 34 Myr. At 30 Myr, material is still present at the inner disc edge, while at 34 Myr, an inner gap of $\sim$20 au has formed due to IPE. A small amount of material is still present within 1 au, but this totals just $\sim2\cdot 10^{-16}$ M$_\odot$, negligible on the scale of a protoplanetary disc.

Protoplanetary discs with large inner gaps are observed, and are typically referred to as transition discs. However, these typically show non-zero accretion rates (e.g. \citet{Owen2012b}). Observations of (non-accreting) weak-lined T Tauri stars with an infrared excess indicating a dust disc, on the other hand, do not reveal gaseous components \citep{Cieza2013,Hardy2015}. This implies that the transition from accreting, gas-rich disc to non-accreting, gas-poor disc is short compared to the disc lifetime. Achieving such a quick transition requires a mechanism that efficiently disperses the disc. \citet{Owen2012} proposed `thermal sweeping', a process where an inner gap allows the host star's X-ray radiation to penetrate into the disc mid-plane. The disc is then dispersed on a few orbital timescales. \citet{Haworth2016} later showed that thermal sweeping is not triggered early enough to result in the observed lack of non-accreting gaseous discs. 

Regardless of the actual process responsible for rapidly dispersing non-accreting discs, the discs in our model are also likely vulnerable to rapid dispersal after accretion ceases. In this way the opening of an inner gap shortens the lifetime of a number of model discs, and lowers the maximum host star mass for which discs reach the fiducial Peter Pan disc age of 50 Myr. For an $\alpha$ viscosity of $10^{-3}$, this maximum mass is lowered from 0.7 M$_\odot$ to 0.6 M$_\odot$, and for an $\alpha$ viscosity of $10^{-4}$ it is lowered from 0.8 M$_\odot$ to 0.55 M$_\odot$.

\subsection{A new population of Peter Pan discs?}

\cite{DeMarchi2017} determined the ages and accretion rates of a large sample of young stars in the 30 Dor Nebula. They found accreting stars with ages up to 50 Myr, presenting a new population of Peter Pan discs. The maximum age of these accreting stars decreased with stellar mass. Within the 1.5-2 M$_\odot$ mass bin, the oldest accreting star had an age of $\sim$25 Myr; $\sim$40 Myr within the 1.1-1.5 M$_\odot$ bin; and $\sim$50 Myr within the 0.5-1.1 M$_\odot$ bin. The oldest stars in the upper two mass bins are considerably older than the corresponding maximum lifetimes from our results ($\sim$20 Myr for 1.5-2 M$_\odot$, and $\sim$30 Myr for 1.1-1.5 M$_\odot$). These ages are for the lowest viscosity (resulting in the longest lifetime) and disregard rapid dispersal after clearing an inner gap. Within the range of 1.1-1.5 M$_\odot$, such rapid dispersal lowers the maximum lifetime to $\sim$20 Myr. 

One explanation could be that the lower metallicity of the LMC enables protoplanetary discs to have greater initial masses. A lower dust content can make the discs less opaque, and so increase the maximum stable mass, as we discussed in Sec. \ref{sec:max_disc_mass}. A larger initial disc mass also helps explain the observed high accretion rates. Many accretion rates are such that if they were sustained over the entire earlier life of the star, an amount of mass comparable to the star itself would have been accreted. 

On the other hand, the oldest stars with masses in excess of 1.1 M$_\odot$ form groups of outliers in the stellar age distribution of their respective mass bins. A bimodal stellar age distribution could be an effect of two phases of star formation. However, the distribution of stellar ages of all masses is not bimodal, implying that the ages of these old stars are potentially overestimated. Disregarding these groups, the oldest star with a mass between 1.1 and 1.5 M$_\odot$ has an age of $\sim$16 Myr, and the oldest star with a mass in the range 1.5-2 M$_\odot$ is $\sim$9 Myr old. These populations would be consistent with our models.

\subsection{Extrapolation of FRIED grid} \label{sec:fried_extrapol}

\begin{figure*}
    \centering
    \includegraphics[width=0.49\linewidth]{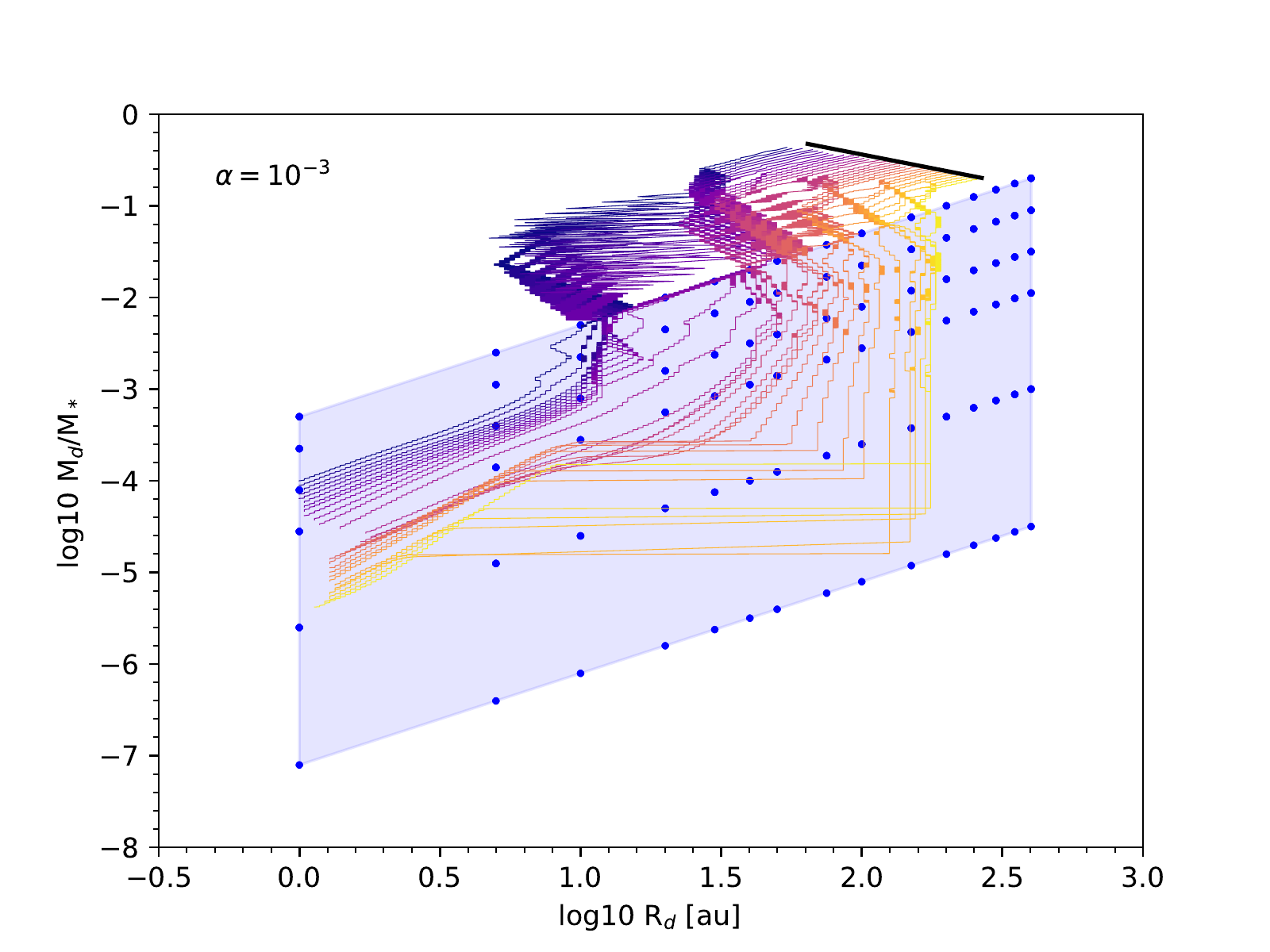}
    \includegraphics[width=0.49\linewidth]{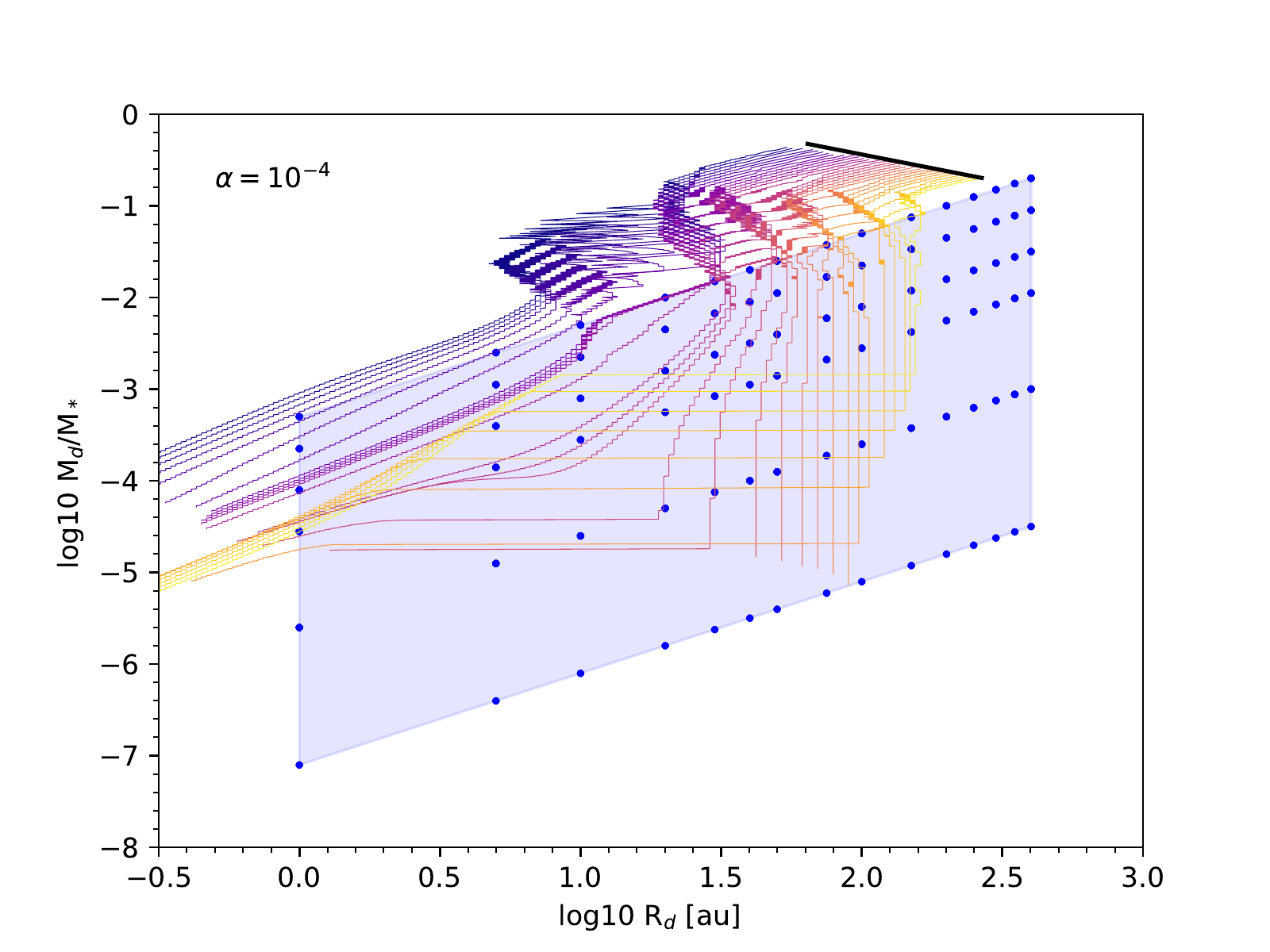}
    \caption{The evolution of the radius and mass ratio of the discs with a viscosity of $10^{-3}$ (left) and $10^{-4}$ (right) under a radiation field of 10 G$_0$. The colour scale indicates initial host star mass, ranging from 0.08 M$_\odot$ (dark purple) to 1.9 M$_\odot$ (yellow). The black solid line indicates the initial disc radii and mass ratios. The blue points correspond to the points of the FRIED grid, and the blue shaded region to the parameter space (convex hull) of these points in log space. Within this region we can interpolate on the FRIED grid, outside we resort to nearest-neighbour extrapolation.}
    \label{fig:fried_path}
\end{figure*}

The initial conditions used in this work are outside of the parameter space of the FRIED grid. (Specifically, outside of the convex hull of the logarithm of the points of the FRIED grid.) Our model discs are initially too compact and too massive, but generally enter the parameter space when they lose mass. When outside the parameter space, we resort to using nearest neighbour extrapolation (in log space) to obtain an EPE rate. In Fig. \ref{fig:fried_path} we show the evolution of our model discs exposed to a radiation field of 10 G$_0$ through parameter space.

We note that the FRIED grid covers a total of four dimensions and is not an entirely regular grid. However, for every radiation field it (mostly) covers the same coordinates for the other three dimensions. It covers a different range of disc masses for every host star mass, but the disc mass fractions are (mostly) the same for every host star mass. The exception is low host star masses ($\le 0.1$ M$_\odot$), at low radiation fields ($\le 1000$ G$_0$), for the two lowest disc mass ratios, which are not on the FRIED grid. However, none of our discs enter this region of parameter space. 

For both viscosities, the discs shrink from the start of the simulation because the EPE rate is larger than the outward mass flux resulting from viscous expansion. All discs initially move parallel to the edge of the FRIED grid, meaning they do not come closer to entering the covered parameter space. When they stop shrinking while still losing mass, most evolve into the parameter space of the FRIED grid. Some discs have large excursions in radius at near constant disc mass outside of the parameter space. These excursions are likely the result of a large difference between the mass loss rate between two neighbouring FRIED grid points. 

The extrapolation of FRIED grid values leads to inaccuracies in the EPE rate of our models that use the FRIED grid. In Fig \ref{fig:mdot}, we saw that the EPE rate tends to increase with disc mass (with other quantities constant). As a consequence, we tend to underestimate the EPE rate for the first phase of the discs' lifetime.

\section{Conclusion}

The discovery of signs of accretion from a gaseous disc around relatively old, low mass stars, prompted us to investigate the survival of such discs under the combined influence of accretion, and internal and external photoevaporation. We did this by running a grid of protoplanetary disc models to estimate the dependence of maximum disc lifetime on host star mass. Our aim was to understand the possibility of the existence of Peter Pan discs around relatively high mass stars. Our conclusions can be summarised as follows:

\begin{enumerate}

    \item The process that limits the maximum possible lifetime of a protoplanetary disc varies with host star mass. For stars less massive than about 0.5 M$_\odot$, external photoevaporation dominates, whereas internal photoevaporation and accretion dominate at higher masses.
    
    \item Discs with mass-radius relations similar to those in young star-forming regions can survive for 50 Myr in low radiation environments for host stars with M$_*$ $\lesssim0.6$ M$_\odot$. 
    
    \item For stars in the range 0.6 M$_\odot$ $\gtrsim$ M$_*$ $\gtrsim 0.8$ M$_\odot$, those discs are limited to lifetimes $\lesssim$50 Myr. We find that before that time, increasingly effective internal photoevaporation chokes accretion and opens up an inner gap which results in rapid disc clearing. 
    
    \item Above that range, discs are limited to lifetimes below 50 Myr because sufficiently efficient internal photoevaporation and accretion disperses them regardless of gap opening. 
    
    \item Peter Pan discs can occur across the entire mass range of M dwarfs. For typical properties of protoplanetary discs and young stars, they are unlikely around stars of K type and earlier.
\end{enumerate}

\section*{Acknowledgements}

MW thanks Margot Leemker for enlightening discussions on observations of transition discs and Nienke van der Marel for a critical read of the manuscript. We also thank the anonymous referee whose insightful comments improved this paper. This paper makes use of the packages {\sc numpy} \citep{Harris2020}, {\sc matplotlib} \citep{Hunter2007}, and {\sc scipy} \citep{Virtanen2020}. 

The simulations were run on an Intel Xeon W-2133 CPU (12 cores, 5 of which were used per run), with every series taking, on average, $\sim$18 hr. Taking into account the 8 series presented here, and a factor 4 for test runs and re-runs, we have a total runtime of $\sim$2880 core hours. Assuming a power usage of $\sim$12 W per core, and 0.649 kWh/kg of CO$_2$ emitted (Dutch norm for grey electricity), this comes to $\sim$50 kg of CO$_2$, comparable to a month of daily commutes. This work was supported by NOVA under project number 10.2.5.12.

%%%%%%%%%%%%%%%%%%%%%%%%%%%%%%%%%%%%%%%%%%%%%%%%%%
\section*{Data Availability}

The code used to obtain the data and the plots presented are available on \url{https://github.com/MJCWilhelm/PeterPanDisks_Public}.



\bibliographystyle{mnras}
\bibliography{PeterPanDiscs} % if your bibtex file is called example.bib

%\begin{thebibliography}{99}
%\end{thebibliography}

%%%%%%%%%%%%%%%%%%%%%%%%%%%%%%%%%%%%%%%%%%%%%%%%%%

%%%%%%%%%%%%%%%%% APPENDICES %%%%%%%%%%%%%%%%%%%%%

\appendix

%%%%%%%%%%%%%%%%%%%%%%%%%%%%%%%%%%%%%%%%%%%%%%%%%%

% Don't change these lines
\bsp	% typesetting comment
\label{lastpage}
\end{document}